\documentclass{raa_twocolumn}            

\usepackage{graphicx,amsmath,amssymb}             

\begin{document}

   \title{From ultraluminous X-ray pulsar to supermassive neutron star
}

   \volnopage{Vol.0 (20xx) No.0, 000--000}      
   \setcounter{page}{1}          

   \author{H. Tong}

   \institute{School of Physics and Materials Science, Guangzhou University, Guangzhou 510006, China;
   {\it tonghao@gzhu.edu.cn}\\
   }

   \date{Received~~2009 month day; accepted~~2009~~month day}

\abstract{The formation of a $2.7\ \rm M_{\odot}$ supermassive neutron star is explored, as the possible companion of PSR J0514--4002E. Magnetars may experience super-Eddington accretion. Observationally they may manifest themselves as ultraluminous X-ray pulsars. We propose that supermassive neutron stars may be formed through ultraluminous X-ray pulsar phase, if the ultraluminous X-ray pulsar phase can last for $10^{5}$--$10^6 \ \rm yr$. The accreted material will also bury the magnetic field of the neutron star. Assuming accretion equilibrium, the final output may be a millisecond supermassive neutron star. In order for the ultraluminous X-ray pulsar phase to last long enough, a magnetic field configuration of the low magnetic field magnetar is required. The mass, magnetic field and rotational evolution of super-Eddington accreting  neutron stars are rather robust against different assumptions, although many of the model details are yet to be determined.
}

   \authorrunning{H. Tong}            
   \titlerunning{From ULX pulsar to supermassive NS}  

   \maketitle

%
%
\section{Introduction}
Recently, a possible supermassive neutron star with a mass of $2.7\ M_{\odot}$ is reported (Barr et al. 2024). According to the observations, the compact star can be either a light black hole or a supermassive neutron star. Irrespective of its nature, it may be formed through the merger of a double neutron star system (Barr et al. 2024). The mass of the compact object ranges from 2.09 to 2.71 $\rm M_{\odot}$ ($95\%$ confidence level, Barr et al. 2024). Here we explore the most extreme possible value and assuming a compact star of mass $2.7 \ \rm M_{\odot}$. Instead of the merger origin proposed in the observational paper, we assume the compact to be a neutron star and investigate how such a supermassive neutron star can be formed.

It is commonly assumed that a neutron star is born with a typical mass about $1.4 \ M_{\odot}$ (Shapiro \& Teukolsky 1983). If the neutron star is in a binary system, during the accretion process, it may accrete about a mass of $\sim 0.2 \ M_{\odot}$ (Zhang et al. 2011). The accreted material will also spin-up the neutron star, to make it a millisecond pulsar (Alpar et al. 1982). Therefore during the accretion process, the mass, magnetic field, and rotational period all evolve with time (Zhang \& Kojima 2006). For a massive neutron star with mass of $2 \ M_{\odot}$ (Demorect et al. 2010; Antoniadis et al. 2013; Cromartie et al. 2020; Fonseca et al. 2021), it may born a little bit more massive or accreted a little bit more material (Tauris et al. 2011; Zhang et al. 2011; Antonoadis et al. 2016). However, the formation of a $2.7 \ M_{\odot}$ neutron star is challenging to previous understandings.

If only $0.2 \ M_{\odot}$ is accreted during the accretion process, then the neutron star will born with a mass of $2.5 \ M_{\odot}$. This is at odd with our previous measured neutron star masses in binaries (Zhang et al. 2011; Antonoadis et al. 2016). Therefore, we prefer to think that the $2.7 \ M_{\odot}$ neutron star is born with a mass about $1.4 \ M_{\odot}$, not atypical with others. The initial mass can also be $1.8 \ \rm M_{\odot}$ if the initial neutron star mass is bimodal (Antonoadis et al. 2016). It is during the accretion process that the neutron star accretes more material than previously thought. And super-Eddington accretion may allow the neutron star to accrete more material.

Among the isolated neutron star population, magnetars may be young neutron stars with magnetic field as high as $10^{14}$-$10^{15} \ \rm G$ (Duncan \& Thompson 1992). This high magnetic field will suppress the scattering cross section between photons and electrons (Herold 1979). This will result in a higher limiting luminosity (Paczynski 1992). The limiting luminosity can be as high as $10^{42} \ \rm erg \ s^{-1}$, whose exact value depends on the magnetic field strength. This explains why magnetars have $10^4$ times super-Eddington luminosity during the pulsating tail of their giant flares (Mereghetti 2008). If the magnetar is born in a binary system, similar physics may also result in a super-Eddington luminosity of accreting neutron star. It is possible that the ultraluminous X-ray (ULX) pulsars may be accreting magnetars (Bachetti et al. 2014; Israel et al. 2017). The X-ray luminosity of ULX pulsar can be as high as $10^{41} \ \rm erg \ s^{-1}$ (Israel et al. 2017). At the same time, they often have a sinusoidal pulse profile. This means that the beaming of ULX pulsars is not significant (Tong \& Wang 2019). Therefore, the accretion rate of ULX pulsars can be 100--1000 times super-Eddington: $10^{-6}$ -- $10^{-5} \ \rm M_{\odot} \ \rm yr^{-1}$. If this super-Eddington accretion phase can last about $\sim 10^5$ -- $10^6 \ \rm yr$, then the neutron star may accrete about $1 \ M_{\odot}$ and become a supermassive neutron star.

However, traditionally, millisecond pulsars are thought to be formed in low mass X-ray binaries. The life time of high mass X-ray binaries is too short to spin-up the neutron star to millisecond period. And the accreted matter in high mass X-ray binary may not have enough angular momentum to form a disk and spin-up the neutron star. Observationally, ULX pulsar may be in intermediate or high mass X-ray binaries (Bachetti et al. 2014; Israel et al. 2017). And theoretical modeling of ULX pulsars found that they may have accrete material in the form a disk (Tong 2015; Tong \& Wang 2019). The only uncertainty is how long can the ULX pulsar phase last? Indirect evidence found that the ULX pulsar phase may last about $10^5 \ \rm yr$ (Belfiore et al. 2020). Indirect observations in other ULX sources indicates a timescale of the super-Eddington accretion about $10^6 \ \rm yr$ (Belfiore et al. 2020). This opens the possibility that the ULX pulsar phase can last for a relative long time.

Linking all this together, a magnetar born in a binary system may accrete at super-Eddington rate and become an ULX pulsar. If the super-Eddington phase last about $10^6 \ \rm yr$, the neutron star may accrete about $1 \ \rm M_{\odot}$ material and become a supermassive neutron star. The massive companion of PSR J0514-4002E may be formed in this way (Barr et al. 2024). Detailed calculations and discussions are given in below.

\section{M, B, P evolution of ULX pulsars}

\subsection{Model calculations}

The ULX pulsar NGC 5907 ULX1 has X-ray luminosity as high as $10^{41} \ \rm erg \ s^{-1}$ (Israel et al. 2017). Possible nebula emission around NGC 5907 ULX1 indicates that the ULX pulsar phase may last about $\sim 10^{5} \ \rm yr$ (Belfiore et al. 2020). Nebulae emission in other ULX sources (it is not known whether they are neutron stars or black holes) indicates that the super-Eddington accretion phase may have a typical duration about $\sim 10^6 \ \rm yr$ (Karret et al. 2017; Belfiore et al. 2020). In the following, we assume that an accretion rate of $10^{-6}$ -- $10^{-5} \ \rm M_{\odot} \ yr^{-1}$ in an ULX pulsar can last about $10^6 \ \rm yr$. The evolution of the neutron star's mass, magnetic field, and rotational period are calculated. According to our calculations, the effect of an accretion rate of $10^{-6} \ \rm M_{\odot} \ yr^{-1}$ lasting for $10^6 \ \rm yr$ is the same as an accretion rate of $10^{-5} \ \rm M_{\odot} \ yr^{-1}$ lasting for $10^5 \ \rm yr$. The mass accretion rate and duration is degenerate. This is especially true for the mass and magnetic field evolution.

The mass evolution of the neutron star is modeled as
\begin{equation}
  M(t) = M_0 + \Delta M(t) = M_0 + \int_0^{t} \dot{M}(t^{\prime}) d t^{\prime} = M_0 + \dot{M} t,
\end{equation}
where $M_0$ is the neutron star mass at the onset of ULX pulsar phase (assumed to $1.4 \ \rm M_{\odot}$), $\Delta M$ is the accreted mass during the ULX pulsar phase. The accreted mass depends on the integral value of mass accretion rate. It is not very sensitive to the detailed dependence of mass accretion rate on time. Therefore, a constant mass accretion rate is assumed.

The magnetic field is modeled in analogy to that of accreting neutron stars in X-ray binaries. After accreting a significant amount of material, the neutron star's magnetic field will decrease significantly. An analytical form of magnetic field evolution due to accretion is
(Shibazaki et al. 1989; Zhang \& Kojima 2006)
\begin{equation}\label{eqn_B_evolution}
  B(t) = \frac{B_0}{1 + \Delta M(t)/m_{\rm B}},
\end{equation}
where $m_{\rm B} \in (10^{-5}, 10^{-3}) \ \rm M_{\odot}$ is a model parameter. $m_{\rm B}=10^{-4} \ \rm M_{\odot}$ is assumed in the following.

It can be seen that both the mass and magnetic field evolution depends only the integral effect of accretion rate, i.e., total accreted mass. The rotational evolution will depend on the detailed time evolution of accretion rate. At the same time, the accretion torque of neutron stars is also uncertain. However, the accretion torque is very effect in spinning-up (accretion phase) or spinning-down (propeller phase), the neutron star may always try to catch its equilibrium period. This will overcome our ignorance of accretion torques.
The magnetospheric radius is defined as the radius where the ram pressure of the accretion flow equals that of the neutron star magnetic field (Shapiro \& Teukolsky 1983; Lai 2014)
\begin{equation}
  R_{\rm m} = \xi \left( \frac{\mu^4}{2 G M \dot{M}^2}  \right)^{1/7},
\end{equation}
where $\xi \sim 1$ is a dimensionless parameter reflecting our ignorance of neutron star magnetospheric physics, $\mu = B R^3$ is the dipole magnetic moment of the neutron star (it is assumed to be the equatorial surface magnetic field $B$ times the neutron star radius $R$ cubed, this is consistent with the definition of pulsar characteristic magnetic field), $M$ is the mass of the neutron star, $\dot{M}$ is the mass accretion rate. The corotation radius is the radius where the Keplerian period equals the neutron star's rotation period (i.e. the synchronous orbit)
\begin{equation}
  R_{\rm co} = \left( \frac{G M}{4\pi^2} \right)^{1/3} P^{2/3},
\end{equation}
where $P$ is the neutron star rotational period. The equilibrium period of accreting neutron stars is defined as the period when the magnetospheric radius equals the corotation radius (if not the neutron star will either spin-up or spin-down)
\begin{eqnarray}\label{eqn_Peq}
  P_{\rm eq}
  &=& 1187\ \xi^{3/2} B_{15}^{6/7} R_6^{18/7} \dot{M}_{17}^{-3/7} M_1^{-5/7} \ \rm s \\
  &=& 1.2\  \xi^{3/2} B_{8}^{6/7} R_6^{18/7} \dot{M}_{17}^{-3/7} M_1^{-5/7} \ \rm ms,
\end{eqnarray}
where $B_{15}$ ($B_8$) are the magnetic field in units of $10^{15} \ \rm G$ ($10^8 \ \rm G$), $R_6$ is the neutron star radius in units of $10^6 \ \rm cm$, $\dot{M}_{17}$ is the accretion rate in units of $10^{17} \ \rm g \ s^{-1}$, $M_1$ is the neutron star mass in units of one solar mass. For an accreting magnetar, the equilibrium period can be longer than 1000 seconds. For a typical magnetic field about $10^8 \ \rm G$, the equilibrium period is about several milliseconds. Usually, the dimensionless parameter $\xi$ is taken as $0.5$ (Ghosh \& Lamb 1979; Shaprio \& Teukolsky 1983). However, the accretion disk may become a thick disk for ULX pulsar (King et al. 2017; Walton et al. 2018). Therefore, the dimensionless parameter $\xi$ is taken as $\xi=1$ during the calculations.

\begin{figure}
\centering
\begin{minipage}{0.5\textwidth}
 \includegraphics[width=0.95\textwidth]{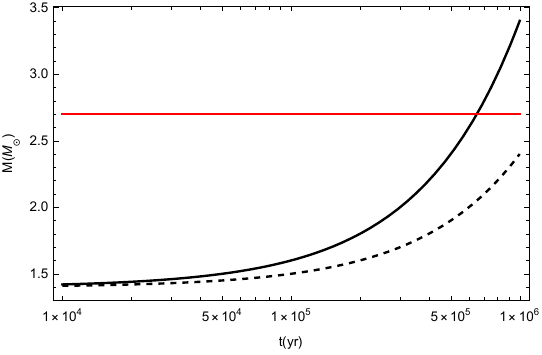}
\end{minipage}
\begin{minipage}{0.5\textwidth}
 \includegraphics[width=0.95\textwidth]{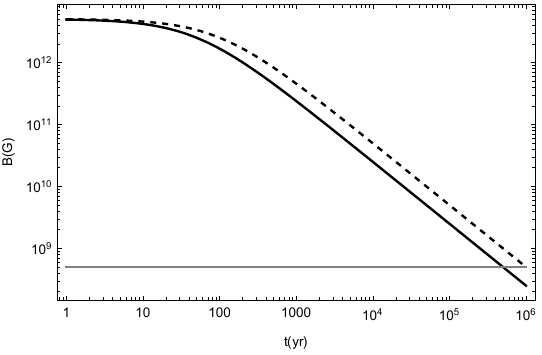}
\end{minipage}
\begin{minipage}{0.5\textwidth}
 \includegraphics[width=0.95\textwidth]{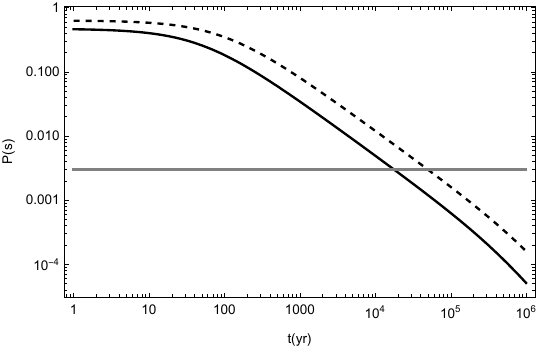}
\end{minipage}
\caption{Mass, magnetic field, and rotational evolution of a neutron star during the ULX pulsar phase.
In each panel, the black solid line and black dashed line are for an accretion rate of $2\times 10^{-6} \ \rm M_{\odot} \ yr^{-1}$ and $10^{-6} \ \rm M_{\odot} \ yr^{-1}$, respectively. In the top panel, the red solid line indicates a mass of $2.7 \rm \ M_{\odot}$. In the middle and bottom panel, the gray solid line indicates a typical magnetic field of $5\times 10^8 \ \rm G$ and a typical period of $3 \ \rm ms$, respectively. The prediction of supermassive neutron is that it should be a millisecond pulsar.}
\label{fig_evolution}
\end{figure}

For a neutron star with mass $M_0 = 1.4\ \rm M_{\odot}$, magnetic field $B_0 = 5\times 10^{12} \ \rm G$, its evolution\footnote{For the case of accreting neutron stars, its initial rotational period is not important, except during the very early stage.} during an ULX pulsar phase is shown in Figure \ref{fig_evolution}. From figure \ref{fig_evolution}, a neutron star experiencing an ULX pulsar phase can accrete a significant amount of mass and become a supermassive neutron star. At the same time, from our previous knowledge of accreting neutron stars, its magnetic field will also decrease significantly. For an initial magnetic field of $10^{12}$--$10^{13} \ \rm G$, its magnetic field will decrease by about $10^{4}$ times after accreting $1 \ \rm M_{\odot}$ of material. For a magnetic field of $10^8$--$10^9 \ \rm G$, the equilibrium period will be several milliseconds if the mass accretion rate is about the Eddington value after the neutron star leaves the ULX pulsar phase.

\subsection{Uncertainties during the modeling}

There are many uncertainties for the long term evolution of an ULX pulsar.
\begin{enumerate}
  \item First of all, how can the ULX pulsar phase maintain if the magnetic field is less than the quantum critical value $4\times 10^{13} \ \rm G$? In the above calculations, the initial magnetic field is already below the quantum critical value. Even if the initial magnetic field is about $10^{15} \ \rm G$, it will be about $10^{13} \ \rm G$ after accreting about $10^{-2} \ \rm M_{\odot}$ of material (according to equation (\ref{eqn_B_evolution})). The answer lies in the existence of low magnetic field magnetars. A low magnetic field magnetar may have a dipole field about $10^{12} \ \rm G$ and multipole field about $10^{14}$--$10^{15} \ \rm G$. The existence of low magnetic field magnetar have been confirmed in case of isolated magnetars (Rea et al. 2010; Zhou et al. 2014). For an accreting low magnetic field magnetar, the low dipole magnetic field is responsible for the interaction between the accretion flow and the neutron star. The magnetic field in the torque related equations is the dipole magnetic field. The multipole field (for example, high magnetic field near the polar cap region) is responsible for the super-Eddington luminosity (Tong 2015; Israel et al. 2017). The scenario of an accreting low magnetic field magnetar is proposed for the first ULX pulsar (Tong 2015) and it is consistent with more observations of ULX pulsars (Israel et al. 2017; Tong \& Wang 2019; Meng et al. 2022).
      Magnetic field reduction during the accretion phase will suppress the dipole magnetic field. At the same time, it may also create
      stronger multipole field (Zhang \& Kojima 2006). This may also allow the existence of accreting low magnetic field magnetars.
  \item For the mass evolution (figure \ref{fig_evolution}, upper panel), there should be a maximum mass for a neutron star. At present the smallest black hole is Cyg X-3 with mass: $M_{\rm BH} = 3.69 \pm 0.41\ \rm M_{\odot} (1\sigma)$ (Tetarenko et al. 2016). A neutron star with $2.7 \ \rm M_{\odot}$ requires a stiff equation of state. According to the equation of state of Lai \& Xu (2009), the maximum mass of a neutron star (actually it is a quark star in Lai \& Xu 2009) can be in the range $M_{\rm NS} \in [3,4] \ \rm M_{\odot}$, which is an amazing result. It seems that we are near the boundary between neutron stars and black holes. When the neutron star in an ultraluminous pulsar phase acquires more mass than its maximum value, it may collapse to a black hole. And the system will become a black hole X-ray binary.
  \item The mass and magnetic field evolution is not very sensitive to the time dependence of accretion rate. They mainly depends on the total accreted materials. However, the period evolution is very sensitive to the accretion rate. For example, the mass accretion rate may switch between a high state with $\dot{M}_{\rm high} \sim 10^{-6} \ \rm M_{\odot} \ yr^{-1}$ and a low state $\dot{M}_{\rm low} \sim 10^{-8} \ \rm M_{\odot} \ yr^{-1}$. This corresponds to the transition between accretion phase (ULX pulsar phase) and propeller phase (Tong 2015; Dall'Osso et al. 2015; Tsykenkov et al. 2016). If the duty cycle of the ULX pulsar phase is about $90\%$, the mass and magnetic field evolution will be unaffected (the required time to reach a certain mass will be slightly longer). During the high state, the neutron star will be spin-up quickly (Carpano et al. 2018). During the low state, the neutron star will try to reach the new equilibrium period (Tong \& Wang 2019). If during the last stage of accretion, the accretion rate is only about $\sim 10^{-8} \ \rm M_{\odot} \ yr^{-1}$ (near the Eddington accretion value), the final neutron star period will be determined by the accretion rate at this time and be around several milliseconds. A supermassive millisecond pulsar may be formed in this way.
  \item For the magnetic field evolution (figure \ref{fig_evolution}, middle panel), there may be a bottom magnetic field (Zhang \& Kojima 2006; Pan et al. 2018). The magnetospheric radius reflects the equilibrium between accretion flow and neutron star magnetic field. When the neutron star magnetic field decreases due to accretion, the magnetospheric radius will also be smaller. This process will continue until the the magnetospheric radius is equal to the neutron star radius: $R_{\rm m} = R$. The magnetospheric radius can not be smaller. This may corresponds to the bottom magnetic field of an accreting neutron star (Zhang \& Kojima 2006; Pan et al. 2018). Numerically, the bottom magnetic field is: $B_{\rm f} = 7\times 10^8 \xi^{-7/4} M_1^{1/4} R_6^{-5/4} \dot{M}_{17}^{1/2} \ {\rm G} \sim (10^8,10^9) \ {\rm G} \propto \dot{M}^{1/2}$. The smaller the mass accretion rate, the smaller the bottom magnetic field (Pan et al. 2018). When neutron star's magnetic field reaches it bottom magnetic field, it may not decrease further even if more materials are accreted.
  \item For the period evolution (figre \ref{fig_evolution}, bottom panel), there may be a minimum period for a neutron star. At present, the fastest spinning neutron star has a rotational period about $P=1.4 \ \rm ms$ (Hessels et al. 2006). The theoretical minim period depends on the neutron star equation of state and the formation process etc (Du et al. 2009). When the corotation radius equals the neutron star radius, this provides an estimation of the limiting period: $P_{\rm min} \sim 0.5 M_1^{-1/2} R_6^{3/2} \ \rm ms$. If the neutron star reaches this limiting period during the accretion process, its period may maintain around this value and spin-up no further.
\end{enumerate}

\section{Discussion and conclusion}

We think that the massive companion of PSR J0514-4002E (Barr et al. 2024) may be a neutron star. We considered the formation of a supermassive neutron star of $2.7 \ \rm M_{\odot}$. The central neutron star may be a low magnetic field magnetar. It can accrete at an super-Eddington rate (i.e. ULX pulsar) and the super-Eddington phase can last for about $10^6 \ \rm yr$. That is: ULX pulsars may result in supermassive neutron stars. The prediction is: supermassive neutron stars are millisecond pulsars. Future surveys of FAST and SKA etc may found such supermassive millisecond pulsars.

At present, the main uncertainties for the long term evolution of ULX pulsars are the duration and duty cycle of the ultraluminous state. The mass and magnetic field evolution are insensitive to the time dependence of accretion rate. They are mainly determined by the total accreted mass. If after leaving the ULX pulsar phase, the accretion rate is maintained around the Eddington value, the equilibrium period will be several milliseconds. Therefore, ULX pulsars will become supermassive millisecond pulsars provided enough mass is accreted. The total accreted mass depends on the duration of the ULX state. Indirect observations indicate that the ULX pulsar state can lasts as long as $10^5 \ \rm yr$ (Belfiore et al. 2020). A duration about $10^6 \ \rm yr$ is also possible according to current observations and theories (Fragos et al. 2015; Wiktorowicz et al. 2015; Karret et al. 2017; Shao et al. 2019; Belfiore et al. 2020). In conclusion, it is possible that ULX pulsar may evolve to supermassive neutron stars. Considering the various uncertainties, more observations are needed in the future.

We have also employed a more detailed magnetic field evolution model (Zhang \& Kojima 2006), instead of equation (\ref{eqn_B_evolution}). The magnetic field evolution model of Zhang \& Kojima (2006) ensures the existence of bottom magnetic field. Except for this point, the difference in the final result is only quantitative. We have also used a more realistic accretion torque model which considers both spin-up torque and spin-down torque (e.g., equaiton (8) in Wang \& Tong 2020). The equilibrium period depends on the magnetic field and mass (equation (\ref{eqn_Peq})). For a decreasing magnetic field and increasing mass, the equilibrium period will decrease with time. Therefore, the neutron star rotational period will always try to catch to the equilibrium period (Wang \& Tong 2020). Again, the difference in the final result is only quantitative. Therefore, the long term evolution of ULX pulsars are rather robust against the different assumptions. Although, many of the model details need to be determined.

ULX pulsars may be intermediate or high mass X-ray binaries (Misra et al. 2020; Abdusalam et al. 2020). At the end of their evolution, the system may become a neutron star white dwarf binary (Shao et al. 2019; Misra et al. 2020; Abdusalam et al. 2020). Current population synthesis have already considered that the accretion disk around ULX pulsar may  become a thick disk. However, an super-Eddington accretion rate due to the effect of strong magnetic field has not be taken into consideration. Even though, a significant amount of information can be drawn for ULX pulsars (Misra et al. 2020; Abdusalam et al. 2020): (1) the ULX phase may have a life time about $\sim 10^6 \ \rm yr$. (2) massive neutron stars are better accretors. A $2.0 \ \rm M_{\odot}$ neutron star can accretes $\ge 0.5 \ \rm M_{\odot}$ masses. These results are consistent with our assumptions and calculations.

Both the merger scenario (Barr et al. 2024) and the ULX pulsar scenario are for the massive companion of PSR J0514-4002E. The formation of the millisecond pulsar itself and the formation of the binary system is beyond the scope of this paper. We note that the binary system of PSR J0514-4002E is found in globular clusters. While some ULX sources are also found in globular clusters (whether they are neutron stars or black hole are not known at present, Dage et al. 2021; Thygesen et al. 2023). This observational aspect is also consistent with the ULX pulsar scenario proposed here. If the massive companion of PSR J0514-4002E is formed through ULX pulsar phase, then the $2.6\ \rm M_{\odot}$ component of GW190814 (Abbott et al. 2020) may also be formed in this way.

\section*{Acknowledgments}
H.Tong would like to thank R.X.Xu for discussions, which inspired the idea of this paper.
This work is supported by National SKA Program of China (No. 2020SKA0120300) and the National Natural Science Foundation of China (NSFC, 12133004).






\label{lastpage}

\end{document}